\documentclass[10pt,conference,a4paper]{IEEEtran}
\IEEEoverridecommandlockouts

\usepackage{mathtools}
\usepackage[utf8x]{inputenc}

\DeclareMathOperator*{\argmax}{arg\,max}

\usepackage{subfigure}
\usepackage{moreverb}
\usepackage{epsfig}
\usepackage{amsmath,amssymb,bm,amsthm,mathrsfs,dsfont}
\usepackage{fancyhdr}
\usepackage{adjustbox,lipsum}
\usepackage[font={small}]{caption}
\usepackage{color,soul}
\usepackage{graphicx}
\usepackage{amsfonts}
\usepackage{cite}
\usepackage{microtype}
\usepackage[colorlinks,bookmarksopen,bookmarksnumbered,citecolor=blue,urlcolor=red]{hyperref}

\usepackage{algorithmicx}
\usepackage{algpseudocode}
\usepackage[norelsize, linesnumbered, ruled, vlined, lined, boxed, commentsnumbered]{algorithm2e}

\allowdisplaybreaks
\usepackage[nodisplayskipstretch]{setspace} 
\begin{document}
\title{ 
AoI-Aware Machine Learning for Constrained Multimodal Sensing-Aided Communications 
}
\author{\IEEEauthorblockN{
 Abolfazl Zakeri\IEEEauthorrefmark{1}, Nhan~Thanh~Nguyen\IEEEauthorrefmark{1},
Ahmed Alkhateeb\IEEEauthorrefmark{2},
 and Markku Juntti\IEEEauthorrefmark{1}}
 \\
 \IEEEauthorrefmark{1}\normalsize 
CWC-RT, University of Oulu, Finland,
 Email: 
\{abolfazl.zakeri,\,nhan.nguyen,\,markku.juntti\}@oulu.fi
\\
\IEEEauthorrefmark{2}The School of Electrical, Computer, and Energy Engineering, Arizona State University, USA, Email: alkhateeb@asu.edu
}
\maketitle
	\begin{abstract}
 Using environmental sensory data can enhance communications beam training and reduce its overhead compared to conventional methods. However, the availability of fresh sensory data during inference may be limited due to sensing constraints or sensor failures, necessitating a realistic model for multimodal sensing.
This paper proposes a joint multimodal sensing and beam prediction framework that operates under a constraint on the average sensing rate, i.e., how often \textit{fresh} sensory data should be obtained. The proposed method combines deep reinforcement learning, i.e., a deep Q-network (DQN), with a neural network (NN)-based beam predictor. The DQN determines the sensing decisions, while the NN predicts the best beam from the codebook.
To capture the effect of limited fresh data during inference, the \textit{age of information} (AoI) is incorporated into the training of both the DQN and the beam predictor. Lyapunov optimization is employed to design a reward function that enforces the average sensing constraint. Simulation results on a real-world dataset show that AoI-aware training improves top-$1$ and top-$3$ inference accuracy by $44.16\%$ and $52.96\%$, respectively, under a strict sensing constraint.
 The performance gain, however, diminishes as the sensing constraint is relaxed.
\end{abstract}
\vspace{-0.5 em}
\section{Introduction} 
Beam management is essential for achieving high performance in millimeter-wave (mmWave) and terahertz (THz) communication systems. Conventional approaches typically rely on beam scanning to identify the best beam, which introduces substantial beam training overhead. Recently, advancements in deep neural networks (DNNs) have enabled learning-based approaches that predict the optimal beam directly from sensory inputs such as position, vision, and LiDAR data~\cite{Ahmed_deepsense,Gerhad_bt,R_Heath_mag,multimodal_wc_mag}; this is broadly referred to as multimodal sensing-aided communications. These methods effectively reduce the beam search space to a small set of high-probability candidates, thereby significantly lowering the beam training overhead.

However, the practical deployment of multimodal sensing-aided beam training methods requires careful consideration of the complexities associated with data acquisition, processing, and the underlying DNN models. Furthermore, the availability of fresh sensory data during inference is often limited by constrained network resources and possible sensor failures. 
Motivated by these observations, this work investigates a beam prediction problem under constrained multimodal sensing.

We propose a dynamic sensing and beam prediction framework for a single user and single base station (BS) subject to a constraint on the average number of multimodal sensing operations, which accounts for resource constraints such as bandwidth, power, and processing overhead.
The proposed framework integrates a deep Q-network (DQN) for dynamic sensor data acquisition and a deep neural network (DNN) for predicting beams using position and camera data.
To model the effect of limited fresh data availability during inference, we incorporate the age of information (AoI) into the training process of both models, where the AoI of each sample represents the time elapsed since the sensory data was captured \cite{Roy_2012,AoI_Monograph_Modiano}. We define a reward function to enforce the sensing constraint using Lyapunov optimization. Simulation results highlight some insightful trade-offs between data freshness and the top-$k$ accuracy of the beam predictor.

\textbf{Related Work:}
Patel and Heath \cite{RHeath_BT_multiuser} proposed a multimodal fusion framework that combines visual and positional sensing with radio-frequency (RF) features for multi-user beamforming. Kim et al. \cite{vision_aid_pos_JSAC_24} exploited vision-aided positioning to improve beam focusing accuracy in THz systems. Work \cite{multmodal_exp_vtc} demonstrated a camera-assisted beam tracking prototype capable of maintaining reliable links under mobility. Arnold et al. \cite{digital_twin} utilized vision-assisted digital twins to enable context-aware beam management, while Vuckovic et al. \cite{multimo_revis} reexamined performance metrics for multimodal beam prediction. 
To reduce the complexity of the underlying DNN, \cite{Ma_KD} proposes the use of knowledge distillation, and \cite{zakeri_DF_KD} presents a data-free knowledge distillation method that enables model compression without requiring access to the original training data. In this context, Park et al. \cite{Walid_ML25} developed a resource-efficient multimodal beam prediction framework that distills knowledge from LiDAR-, radar-, RGB-, and GPS-based models into a lightweight radar-only network. Collectively, these studies demonstrate that incorporating visual and spatial modalities can significantly improve beam alignment efficiency and robustness compared to RF-only schemes. Nonetheless, the impact of sensing constraints and the availability of fresh data during inference remain largely unexplored.

The most closely related works to this paper are \cite{Ahmed_vision} and \cite{zakeri_drl_sen}. In \cite{Ahmed_vision}, the authors developed a multimodal machine-learning framework that integrates vision and GPS (position) data with mmWave beam training measurements to predict optimal beams in real vehicular communications scenarios. Building on this, our work extends \cite{Ahmed_vision} to a \textit{constrained} multimodal sensing framework and further extends our previous study \cite{zakeri_drl_sen} to \textit{multimodal} sensing that incorporates AoI into the beam predictor training, thereby improving beam prediction performance under strictly limited sensing conditions.
\section{System Model and Problem Formulation}\label{system_model}
We consider a downlink mmWave communications system consisting of a BS and a single-antenna mobile user equipment (UE). The BS is equipped with $N$ antennas and an RGB camera. 
The BS adopts a predefined analog beamforming codebook $\mathcal{F} = \{\mathbf{f}_1, \dots, \mathbf{f}_M\}$ for signal transmission, where $\mathbf{f}_m\in\mathbb{C}^{N\times 1}$ with ${\|\mathbf{f}_m\|_{2}^{2} = 1}$, ${m=1,2,\dots, M}$.
\\\indent 
Denote by $\mathbf{h}(t)\in\mathbb{C}^{N\times 1}$ the channel between the BS and the UE at time slot $t$. Furthermore, let ${x}(t)\in\mathbb{C}$ be the transmit data symbol from the BS to the UE at slot $t$, with $\mathbb{E}\{|{{x}(t)}|^2\}=P$, where $P$ is the transmit power.
Suppose the beamforming vector $\mathbf{f}(t)\in\mathcal{F}$ is chosen at slot $t$, the received signal at the UE is then given by 
\begin{align}
    y(t)= \mathbf{h}^{\textsf{H}}(t) \mathbf{f}(t){s}(t) + n(t).
\end{align}
Here  $n(t)\in \Bbb{C}$ is additive white Gaussian noise (AWGN) following the distribution $\mathcal{CN}(0,\sigma^2)$, with $\sigma^2$ denoting the noise variance at the UE.
\\\indent 
In the conventional beam scanning methods, best beam~$m$ at time $t$ is chosen from the codebook to maximize the received signal-to-noise ratio (SNR) $\dfrac{|\mathbf{h}^{\mathsf{H}}(t)\mathbf{f}_m(t)|^2}{\sigma^2}$.
 This requires the channel state information or certain search-based beam training procedures. The former is typically hard to obtain in mmWave/THz systems, and the latter incurs excessive beam training overhead and latency. 
 
To address these challenges, the primary goal of this paper is to develop a multimodal sensing-aided beam prediction that utilizes visual and position data to predict the optimal beam to maximize the received SNR. 
Particularly, we will utilize RGB images captured by the camera installed at the BS and
 the real-time position information of the UE. 
 Let $\mathbf{p}(t)\in\mathbb{R}^2$ be the two-dimensional position vector
(latitude and longitude) of the UE. Furthermore, let $\mathbf{X}(t)\in\mathbb{R}^{W\times H\times C}$ be the RGB image captured at slot $t$, where  $W$, $H$, and $C$ are respectively the width,
height, and the number of color channels of the~image.

Existing works on multimodal sensing-aided beam prediction assume the availability of the \textit{freshest data}, i.e., $\{\mathbf{p}(t),\mathbf{X}(t)\}$, (during inference) to predict the best beam at time slot~$t$. This, however,  necessitates continuous sensing, bandwidth, and power, and in general, excessive resource consumption and additional complexity. 
Thus, there is a trade-off between the sensing resources required to obtain fresh multimodal data (and the subsequent processing complexity) and the resulting beam-prediction performance.

To capture the freshness of sensory data used for multimodal beam prediction, we employ AoI, defined as the time elapsed since the most recent sample (position+RGB) was captured. To formulate the limited-sensing beam prediction problem, let $\alpha(t)\in\{0,1\}$\footnote{In general, this can be defined separately for each sensing modality.} denote the sensing decision at time slot~$t$, where $\alpha(t)=1$ indicates that sensing is performed and ${\alpha(t)=0}$ otherwise. More precisely, ${\alpha(t)=1}$ means that the current sensing data at time $t$ are acquired and made available for beam prediction. Given the sensing decision, AoI evolves as
\begin{align}\label{eq_agedynamic}
   \delta(t+1) = 
   \begin{cases}
     0, & \text{if } \alpha(t)=1, \\
     \delta(t)+1, & \text{if } \alpha(t)=0.
   \end{cases}
\end{align}

\begin{figure}[t!]
\subfigure[Sample 1300] 
{
\includegraphics[width=0.22\textwidth]{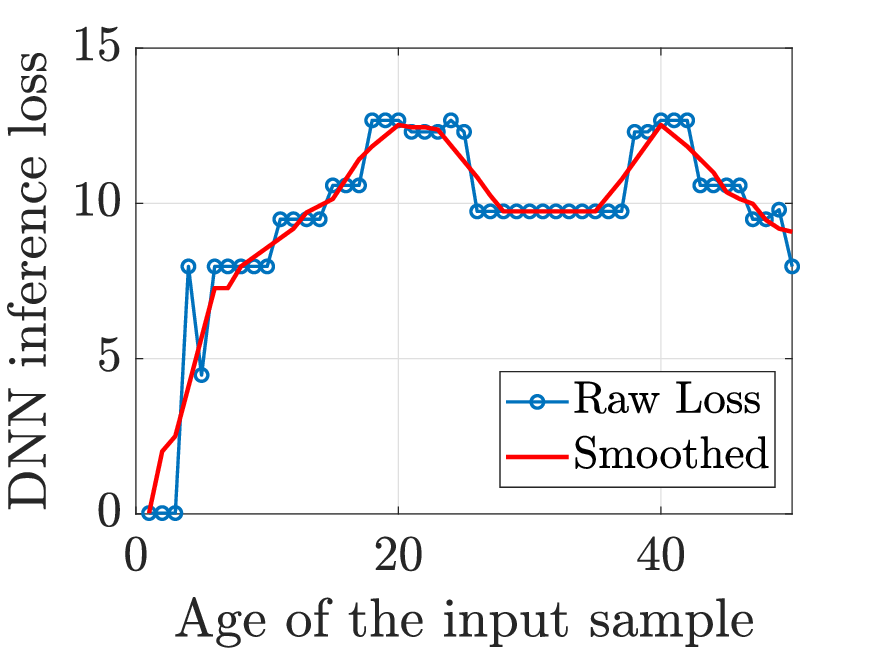}
\label{fig_gnrloss_Top1}
}
\subfigure[Sample 250]{
\includegraphics[width=0.22\textwidth]{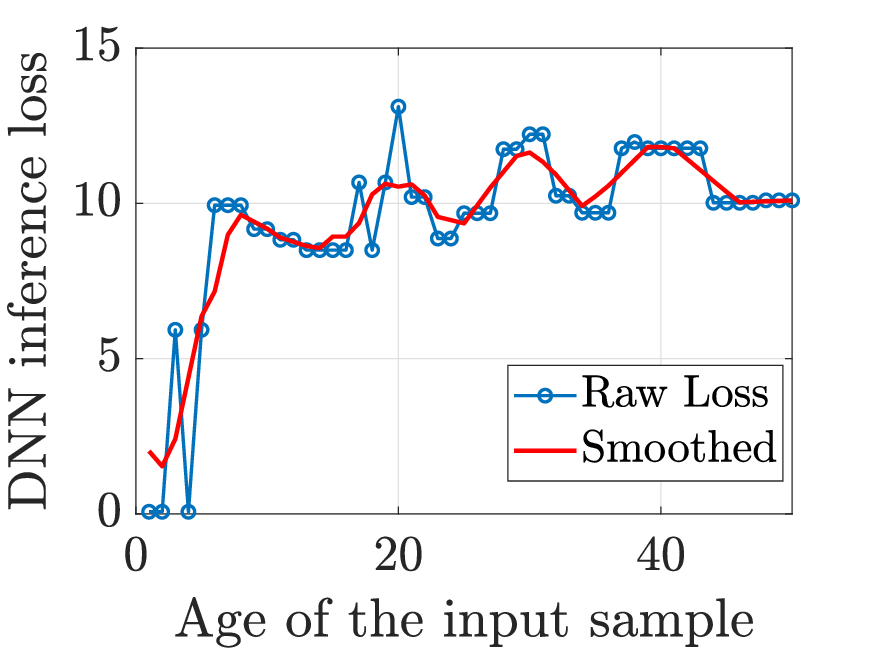 }
\label{fig_gnrloss_Top5}
}
\vspace{-0.2 em}
\caption{Cross-entropy inference loss versus input sample age (freshness) for the image + position data in Scenario 5 of the DeepSense dataset \cite{Ahmed_deepsense}. Increasing age on the x-axis indicates that the same (older) sample is repeatedly fed into the (pretrained) model starting from sample at $t=1300$ (a) and $t=250$ (b). }
    \label{fig_imgpos_infloss_age}
    \vspace{-1 em}
\end{figure}

\textit{\textbf{Inference error/loss}:} Let $\mathcal{X}(t)$ denote the sensory data acquired and made available for the beam prediction at time $t$. 
Using vision and position multimodal data, ${\mathcal{X}(t) := \{\mathbf{p}(t), \mathbf{X}(t)\}}$. Then, by definition, $\mathcal{X}(t-\delta(t))$ corresponds to the data sampled in $\delta(t)$ time slots earlier, i.e.,  $\mathcal{X}(t-\delta(t)) = \{\mathbf{p}(t-\delta(t)), \mathbf{X}(t-\delta(t))\}$. 
 Since continuous sensing is not always feasible, as dictated by the sensing decision $\alpha(t)$, at each time slot $t$ in the inference, the (multimodal) DNN-based beam predictor utilizes $\mathcal{X}(t-\delta(t))$ to predict the optimal beam index $m^*(t)$. The performance of this prediction is measured via a bounded loss function $f(t)$, which we specify explicitly later in Sec.~\ref{sec_beam_pridictor}. 
 Notably, $f(\cdot)$ is a function of the sensing decision $\alpha(t)$, and hence $\delta(t)$, through the data sample $\mathcal{X}(t-\delta(t))$ and the selected beam $m(t)$. However, as the beam prediction is data-driven, the loss function $f(\cdot)$ is hard to explicitly characterize as a function of $\alpha(t)$.
For instance, in Fig.~\ref{fig_imgpos_infloss_age}, we illustrate the cross-entropy log loss as a function of the age of the predictor image and position input samples for $\delta(t)=1,\dots,50$. Notably, as can be seen from this illustration, the inference loss is generally a nontrivial and \textit{non-monotonic} function of~AoI. For noticeably large values of age (e.g., $\delta(t) > 20$), however, the inference loss remains high, suggesting that reducing the loss requires acquiring a fresher sample. 
\\\indent
We now formulate our constrained multimodal sensing-aided beam prediction problems, as a tradeoff between the sensing budget, i.e., sensing rate, and the beam prediction loss, as follows: 
\begin{subequations}
\label{op_1}
\begin{align}
\underset{\{m(t), \alpha(t)\}_{t=1,2,\ldots}}{\mbox{minimize}}~~~ & 
\lim_{T\rightarrow \infty } \frac{1}{T} \sum_{t=1}^{T} 
\mathbb{E}\!\left\{ f(t)  \right\} \label{eq_obj_fun} \\
\mbox{subject to}~~~~~~~~ &  
\lim_{T\rightarrow \infty } \frac{1}{T} \sum_{t=1}^{T} 
\mathbb{E}\!\left\{ \alpha(t)\right\} \leq \alpha^{\max}, \label{eq_cons_sensing}
\end{align}
\end{subequations}
where $m(t)\in\{1,\dots,M\}$ is the beam index selection variable, $\alpha(t)\in\{0,1\}$ is the sensing decision, and $\alpha^{\max}\in(0,1]$ denotes the (normalized) sensing (cost) budget. Moreover, the expectation is with respect to the possibly randomized determination of the variables. Constraint \eqref{eq_cons_sensing} represents a sensing rate/budget limit; the constraint generally reflects practical resource limitations, including bandwidth, power, and other factors associated with sensing execution, (pre)-processing, and transmission of the sensed data.\footnote{This model captures the fundamental trade-off between data availability and beam prediction performance. Extension to a distributed sensor system with a physical communication channel between the predictor and sensors is left for future work.} 
We assume the available information for solving problem \eqref{op_1} at each time $t$ is AoI and the corresponding sensory data $\{\mathbf{p}(t-\delta(t)), \mathbf{X}(t-\delta(t))\}$. 
It is worth noting that, in general, utilizing all previously taken samples may further improve performance due to possibly temporal correlations in the sensory data. However, doing so would naturally increase both the complexity of the beam predictor and the sensing decision-making design. 

\begin{figure}
    \centering
    \includegraphics[width=0.99\linewidth]{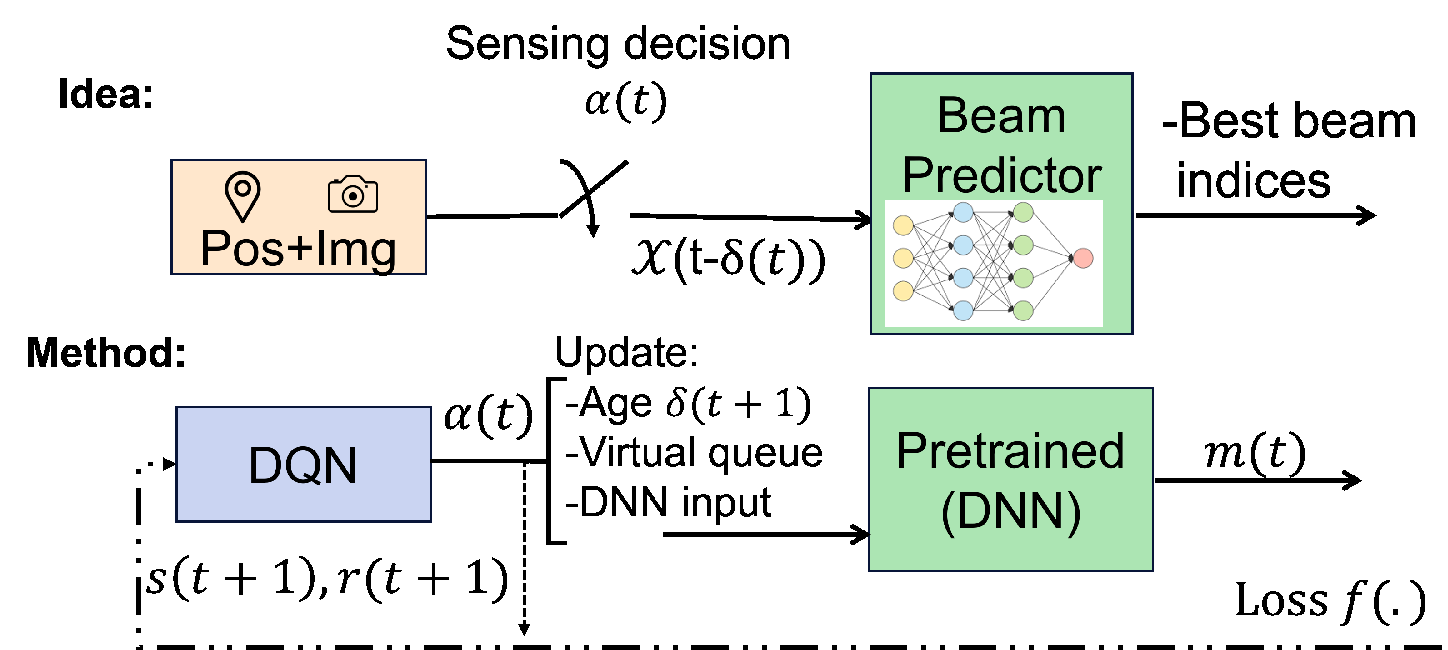}
    \caption{Schematic of the proposed (sequential) joint sensing and beam prediction.}
    \label{fig_method_seq}
    \vspace{-1 em}
\end{figure}

 \section{Proposed Solution to Problem \eqref{op_1}}
In this section, we develop a heuristic method to solve problem \eqref{op_1}, as schematically illustrated in Fig.~\ref{fig_method_seq}.
Our method \textit{sequentially} determines the sensing decision $\alpha(t)$ using a DQN and the beamforming decision $m(t)$ using a separate DNN predictor.
To obtain the beam prediction, the DNN-based predictor uses the available sensory data of position and RGB at each time slot $t$.
We explain in Sec.~\ref{sec_beam_pridictor} how this predictor is trained to determine the beam index decisions. Once trained, during inference (test), it uses the available data dictated by the sensing decisions $\alpha(t)$. Therefore, constraint~\eqref{eq_cons_sensing} is (only) involved in the sensing decisions. As such, during the training of the beam predictor module, we do not account for this constraint.
It should also be noted that problem \eqref{op_1} cannot be solved in a fully data-driven manner, as there are no desired or labeled outputs corresponding to the optimal decisions for $\alpha(t)$. Therefore, we propose a reinforcement learning method to learn the optimal sensing decisions in~Sec.~\ref{sec_drl}.
\subsection{Training the Beam Predictor}\label{sec_beam_pridictor}
For training the beam predictor, multimodal data comprising position and image features, along with their corresponding true labels $y(t)$, are used for each time index $t$. To incorporate the AoI into model training, an augmented dataset is constructed based on the original samples. Specifically, given any AoI value $\delta(t)$ and time index $t$, new data samples are generated as $(\mathcal{X}(t - \delta(t)), \delta(t); y(t))$, whereas the original dataset consists of pairs $(\mathcal{X}(t); y(t))$.
This augmentation strategy is designed to reflect real-world inference scenarios, where the availability of fresh samples is constrained by the sensing limitation in~\eqref{eq_cons_sensing}.
Including AoI as a feature and training on repeated samples with different ages enables the model to learn the relevance of outdated samples and appropriately associate them with the target labels.
It is also important to note that AoI values must be bounded to maintain a finite dataset size. Beyond a certain threshold, larger AoI values have diminishing or even degrading effects on predictor performance as the corresponding data become too outdated.
Sec.~\ref{sec_numres} analyzes the impact of incorporating AoI into model training on \textit{inference} performance under the sensing constraint. 

To perform beam prediction, we adopt the same NN architecture presented in~\cite[Fig.~2]{Ahmed_vision}, with the modification that the AoI value is concatenated with the corresponding image features and 2D-normalized position data for the model’s input.
For details of the NN architecture, we refer the reader to~\cite[Sec.~III-B]{Ahmed_vision}.

\subsection{Deep Reinforcement Learning for Dynamic Sensing}\label{sec_drl}
Given the \textit{pretrained} beam predictor, the beam selection variable $m(t)$ is no longer an optimization variable, and the task reduces to solving problem \eqref{op_1} with respect to the sensing decision $\alpha(t)$.
This optimization can be formulated as a constrained deep reinforcement learning (DRL) problem.

In the standard Lagrangian method, a two-step iteration updates the Lagrange multiplier after solving the resulting unconstrained Markov decision process (MDP), which can be computationally expensive.
Alternatively, Lyapunov-based optimization~\cite{Neely_Sch} provides a simpler way to handle constraint~\eqref{eq_cons_sensing} by introducing a virtual queue for constraint violations. Inspired by the drift-plus-penalty framework~\cite{Neely_Sch, Zakeri_Journal_Relay}, we use an upper bound of the drift-plus-penalty term as the DQN’s immediate reward, to enforce the average constraint \eqref{eq_cons_sensing}. Next, we begin with some key definitions.

Let $Q(t)$ denote the virtual queue associated with constraint~\eqref{eq_cons_sensing} at slot $t$, whose evolution is given by 
\begin{equation}\label{eq_virtualQueue}
    Q(t+1) = \max \Big[ Q(t) + \alpha(t) - \alpha^{\max} ,\, 0 \Big].
\end{equation}
The dynamics of $Q(t)$ above can be seen as a queue with arrival rate of $\alpha(t)$ and a service rate $\alpha^{\max}$. Constraint \eqref{eq_cons_sensing} is satisfied if this virtual queue becomes strongly stable, i.e.,  
$
\limsup_{T\rightarrow \infty} \frac{1}{T}\sum_{t=1}^T \mathbb{E}\{Q(t)\} < \infty
$ \cite[Ch. 2]{Neely_Sch}.

Define a scalar measure of the queue congestion (or queue size) by $L(Q(t)):= \dfrac{1}{2}Q(t)^2$. Then define $\Delta(Q(t))$ as the conditional Lyapunov drift for slot $t$: \cite[Eq. 3.13]{Neely_Sch}
\begin{equation}\label{eq_drift}
    \Delta(Q(t)):=\mathbb{E}\!\left[ L(Q(t+1)) - L(Q(t)) \,\big|\, Q(t) \right].
\end{equation}
where the expectation is with respect to the (possibly) randomized action selection of $\alpha$. Note that this expectation is in general with respect to system randomness, e.g., probabilistic dynamics of the age of information, but in our system model, there is no such randomness given a sensing decision. 

Leveraging the fact that for any $c \ge 0, b \ge 0, A \ge 0$, we have \cite[p. 33]{Neely_Sch}
\begin{align}
    (\max[c-b,0]+A)^2 \le c^2+A^2+b^2+2c(A-b).
\end{align}
One can derive the following upper-bound for $\Delta(Q(t))$:
\setlength{\abovedisplayskip}{5pt}
\setlength{\belowdisplayskip}{5pt}
\begin{equation}
    \Delta(Q(t)) \le C + Q(t) \Bbb{E}\{\alpha(t)-\alpha^{\max}\,|\,Q(t)\},
\end{equation}
where $C$ is a positive constant.

Now, assuming that the beam selection variable $m(t)$ is given, we consider the upper bound above, drop the constant terms and cast problem \eqref{op_1} as follows:
\begin{align}\label{eq_op_drl}
 \underset{\{\alpha(t)\}_{t=1,2,\ldots}}{\mbox{maximize}}~~\lim_{T\rightarrow \infty} \dfrac{1}{T} \sum_{t=1}^{T}   \mathbb{E}\{Q(t)\alpha(t) + Vf\left(t \right)\}.
\end{align}
where $V$ is a non-negative parameter to desirably adjust a trade-off between the size of the virtual queue and the beam prediction loss. 

We employ the DQN approach to solve (the discounted version of) the above problem. The action, state, and reward components are defined as follows.
\\ $\bullet$ \textit{Action:} The action is the sensing decision $\alpha(t)$, i.e., ${a(t)\in\{0,1\}}$.
\\
$\bullet$ \textit{State:} The state provides information for the sensing decisions. We compose the state as the AoI, the virtual queue, and the most recent sensory data, i.e.,
\begin{align}
    s(t):= (\delta(t),Q(t))).
\end{align}
$\bullet$ \textit{Reward:} 
The reward function represents the immediate objective of problem~\eqref{eq_op_drl}, i.e.,
\begin{align}\label{eq_reward}
    r(s(t), a(t)) := -\big[\, V f(t) + Q(t)a(t) \,\big],
\end{align}
where recall that $f(t)$ denotes the beam prediction loss. We next specify our choice of the function $f(\cdot)$ in accordance with the goal that the beam predictor should assign a higher softmax probability to the optimal beam index. This choice also influences the DQN action, which determines the input data provided to the DNN predictor. 
Let ${p_m(t) := \Pr\{m = y(t)\}}$  be the softmax output of the DNN, given by
\begin{align}
    p_m(t) = \mathsf{softmax}(z_m(t)) 
    := \frac{e^{z_m(t)}}{\sum_{j=1}^M e^{z_j(t)}}, 
    \quad m = 1,\dots,M,
\end{align}
where $z_m(t)$ is the $m$-th logit, i.e., the raw model output. 
Ideally, $p_m(t)$ should be a delta function, i.e., ${p_m(t) = \delta(m - y(t))}$. 
Therefore, the beam prediction loss is defined via the cross-entropy as:
\begin{align}\label{eq_entropyloss}
    f(t) := -\sum_{m=1}^M \delta(m - y(t)) \log(p_m(t)) 
    = -\log\big(p_{y(t)}(t)\big).
\end{align}
where recall $y(t)$ is the true (optimal) label.
\noindent
Note that other choices for $f(t)$ are also possible; for instance, 
${ f(t) = \mathds{1}_{\{\argmax_m p_m(t) = y(t)\}} }$. 
However, our method is not restricted to this particular form and remains applicable for any \textit{bounded} function.

Briefly, the DQN employs a NN to approximate the optimal action-value function, i.e., the Q-function~\cite{Deep_Learning_Nature}:
\begin{align}
    Q(s,a) := \mathbb{E} \!\left[ \sum_{t=0}^{\infty} \gamma^t r(s(t), a(t)) \,\big|\, s(0)=s,\, a(0)=a \right],
\end{align}
where $\gamma$ is the discount factor. 
The sensing action is then greedily selected for each state $s$~as
\begin{align}
    \pi(a|s) := \arg\max_{a} Q^{*}(s,a).
\end{align}
During DQN training, the pretrained beam predictor is used to determine the beam selection action $m(t)$, which in turn allows the computation of the reward.

In summary, the proposed constrained sensing and beam prediction consists of three main stages:  
(1) augment a training dataset by using image–position samples with their true labels, and age values,
(2) training a DNN for beam prediction using this augmented dataset that includes AoI values, and  
(3) training a DQN to learn a sensing policy that maximizes beam prediction accuracy while ensuring virtual queue stability to satisfy the sensing budget constraint.  
These steps are outlined in Alg.~\ref{alg:dnn_dqn_training}.
\begin{algorithm}[t]
\caption{\small AoI-aware Constrained Sensing and Beam Prediction }
\label{alg:dnn_dqn_training}
\small
\tcc{Initialization}
Set parameters: sensing limit $\alpha^{\max}$, $V$, get image-position data with labels (as a dataset), and initialize DNN and DQN parameters\;

\tcc{Step (1): Augment a (new) dataset}

Replicate each sample at slot $t$ for every value of $\delta(t)$ with updated labels from $t+\delta(t)$

\tcc{Step (1): Train a DNN for beam prediction}

Set the model as in \cite[Fig. 2]{Ahmed_vision}

Concatenate the age value to the input layer

\For{each epoch}{
    Sample a mini-batch from the augmented dataset\;
    Compute predictions and cross-entropy loss\;
    Update DNN parameters via backpropagation\;
    Save the trained model for the next step\;
}

\tcc{Step (3): Train DQN for sensing decisions}
\For{each episode}{
    Initialize environment and generate a random state $s(0)$\;
    \For{each time step $t=1,2,\dots,T$}{
        Choose action $a(t)$ using the $\epsilon$-greedy policy\;
        Execute $a(t)$: update $Q(t+1)$ by \eqref{eq_virtualQueue} and age $\theta(t+1)$ by \eqref{eq_agedynamic}, then obtain $s(t+1)$\;
        Use $\mathcal{X}(t-\theta(t+1))$ and the trained DNN (Step 2) to compute the cross-entropy loss in \eqref{eq_entropyloss}\;
        Compute the reward function in \eqref{eq_reward}\;
        Store transition $(s(t), a(t), r(t), s(t+1))$ in replay memory\;
        Sample a mini-batch from replay memory and update DQN parameters via Q-learning\;
        Set $s(t) \leftarrow s(t+1)$\;
    }
}

\tcc{Output}
Return the trained DNN and DQN for real-time inference~(Fig.~\ref{fig_method_seq})\;
\end{algorithm}
\section{Numerical Results}\label{sec_numres}
Here we present simulation results for the position and image modality.\footnote{The source code is available on \href{https://github.com/AZakeri94/AoI-aware-ML-for-position-image-aided-beam-prediction-.git}{GitHub}.}
We first train the beam predictor offline using two different dataset: (1) is the original Scenario~5 in DeepSense dataset \cite{Ahmed_deepsense}, and (2) the dataset of Scenario 5 but with the modified samples, i.e., for every sample in the dataset, we augment new rows $(\mathcal{X}(t-\delta(t),\delta(t))$ for all values of $\delta(t)\in\{1,\dots,N\}$, where $N $ is a finite number. The model is adopted from \cite{Ahmed_vision} with the modification that the input also includes the age values. The model is trained for $15$ epochs with learning rate $10^{-3}$ and batch size $32$. Once it is trained, the model is saved for training the DQN module in Fig.~\ref{fig_method_seq}. Performance is evaluated using top-$k$ accuracy, defined as the probability that the optimal beam lies within the top-$k$ predicted beams.

\textit{Training DQN:} We implement DQN with a three-layer fully connected network with 64 neurons in each hidden layer. The input dimension matches the DQN state size, and the output dimension corresponds to the action space, which predicts $2$ Q-values corresponding to each action. Training is performed with a discount factor of $0.99$, a learning rate of $0.001$ using the Adam optimizer, a batch size of 64, and a replay memory of 50{,}000. The model is trained for 100 epochs, each consisting of 300 iterations. In the train of DQN, we use the pretrained beam predictor neural network to compute the reward function~\eqref{eq_reward}.

\begin{figure}[t!]
\subfigure[Average number of sensing $\bar{\alpha}(t)$] 
{
\includegraphics[width=0.22\textwidth]{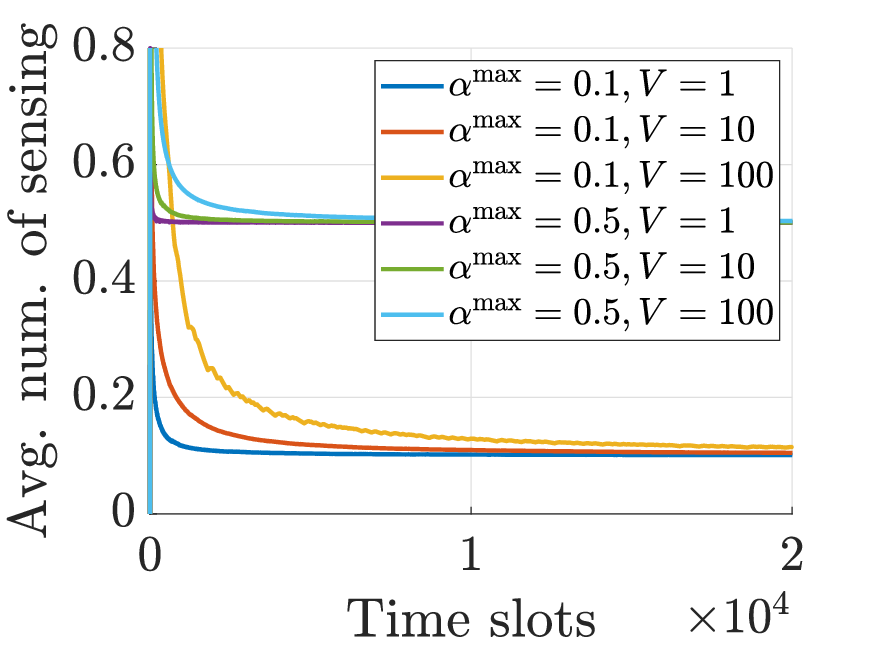}
\label{fig_avg_alpha}
}
\subfigure[Average virtual queue size]{
\includegraphics[width=0.22\textwidth]{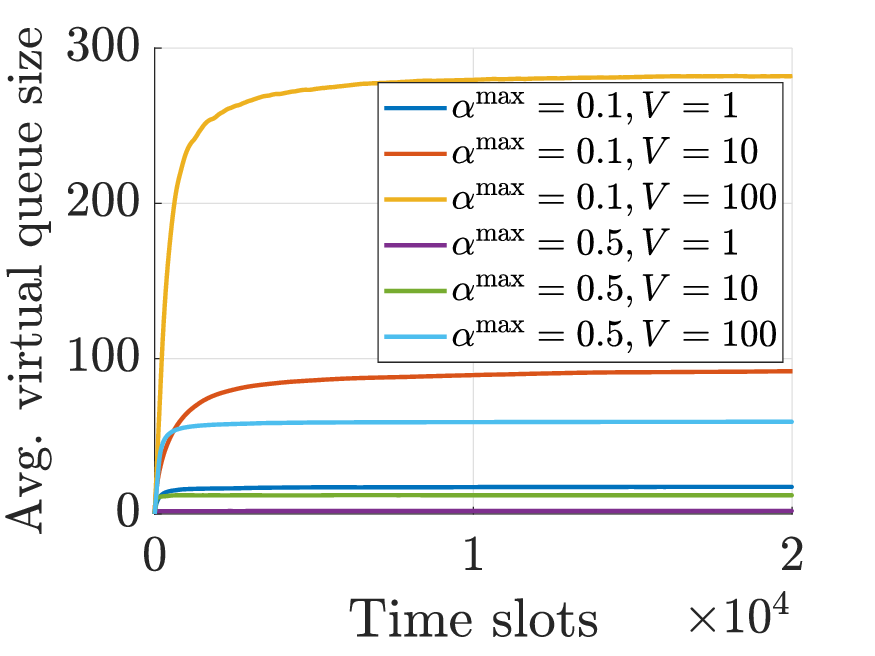}
\label{fig_avg_queue}
}
\caption{Satisfaction of the average constraint \eqref{eq_cons_sensing} by the proposed DQN algorithm for different values of parameter $V$. The average number of sensing (a) and average virtual queue size $Q(t)$ (b) vs. time slots at the inference. }
    \label{fig_cons}
\end{figure}

We first present in Fig.~\ref{fig_cons} that the proposed DQN-based sensing method, with the designed reward function, successfully satisfies the average constraint in \eqref{eq_cons_sensing} for different values of the sensing budget $\alpha^{\max}$ and the control parameter~$V$. As illustrated in Fig.~\ref{fig_avg_queue}, the virtual queue under our method remains strongly stable for any reasonable choice of $V$, i.e.,
$
\limsup_{T \to \infty} \frac{1}{T}\sum_{t=1}^T \mathbb{E}\{Q(t)\} < \infty,
$
which theoretically guarantees the satisfaction of the constraint in \eqref{eq_cons_sensing} \cite[Ch. 2]{Neely_Sch}.
Nevertheless, it is observed that larger values of $V$ lead to slower convergence toward the constraint limit $\alpha^{\max}$ and cause an initial buildup of the queue during early iterations.
 
\begin{figure}[t!]
\subfigure[Top-$1$ Accuracy] 
{
\includegraphics[width=0.22\textwidth]{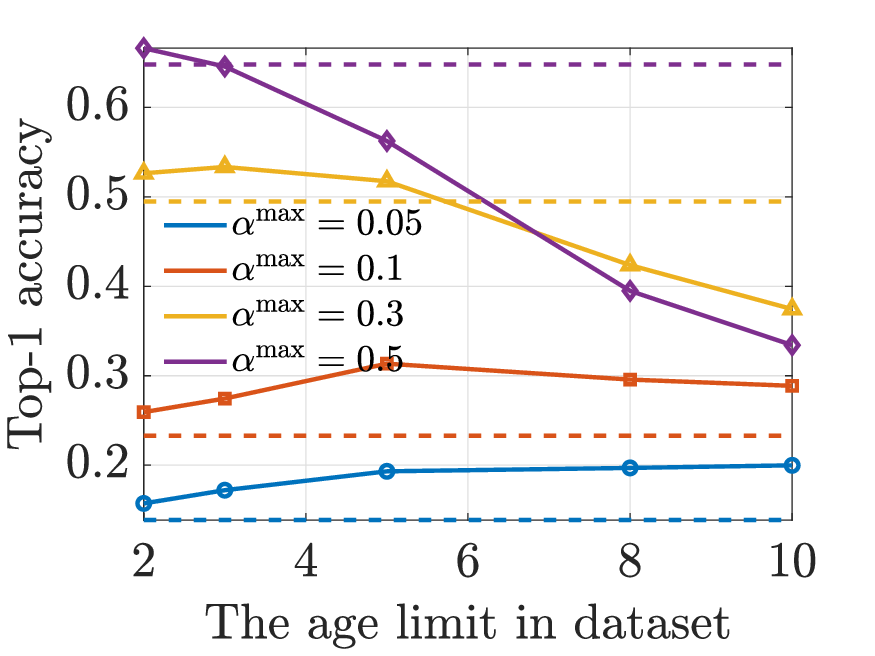}
\label{fig_age_limit_top1}
}
\subfigure[Top-$3$ Accuracy]{
\includegraphics[width=0.22\textwidth]{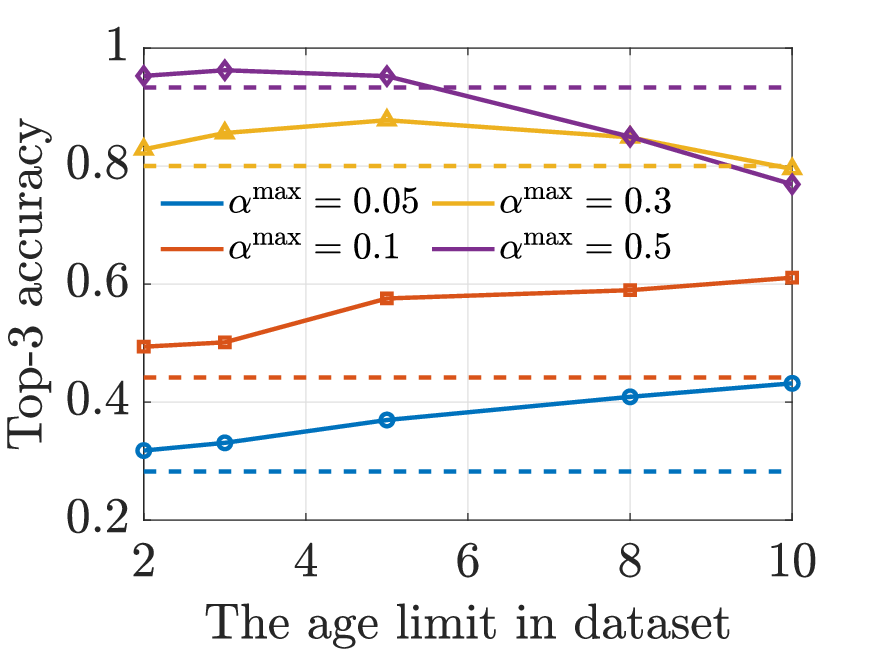}
\label{fig_age_limit_top3}
}
\caption{Top-$k$ inference accuracies versus age limit in the augmented dataset for different sensing budgets, $\alpha^{\max}$, under randomized sensing decisions.}
    \label{fig_age_limit}
\end{figure}

\textit{Impact of AoI in beam predictor performance:} We train the beam predictor (i.e., the DNN module in Fig.~\ref{fig_method_seq}) using an augmented dataset with age values $\delta \in {1, 2, \dots, N}$. Each data sample at time $t$ is replicated $N$ times with updated labels from time $t+\delta(t)$ and an additional column including the age values.
Figure~\ref{fig_age_limit} illustrates the impact of the age limit $N$ on the inference performance under the sensing constraint in \eqref{eq_cons_sensing}. To isolate the effect of the beam predictor, sensing decisions are randomized while satisfying \eqref{eq_cons_sensing}. The dashed line represents the baseline model trained on the original dataset without augmentation.
Larger $N$ values improve performance for small sensing budgets (e.g., $\alpha^{\max} \leq 0.3$), but the gain diminishes as the budget becomes more relaxed (i.e., $\alpha^{\max} \rightarrow 1$). The performance improvement from age augmentation is more evident in the Top-$3$ accuracy, while the inclusion of many outdated samples increases training difficulty, particularly for Top-$1$ accuracy.
Overall, $N$ should be carefully selected based on the sensing budget. A practical choice is $N = \lceil 1 / \alpha^{\max} \rceil$, since $1 / \alpha^{\max}$ approximates the average reuse count of outdated samples. Hence, when $\alpha^{\max}$ is close to 1 (e.g., $\alpha^{\max} > 0.8$), age-based dataset augmentation may not be necessary.

\textit{Performance comparisons:} We now present an inference performance comparison of the proposed method, DQN sensing combined with a predictor DNN trained on the age-augmented dataset (referred to as “DQN with age predictor”). For benchmarking, we consider four cases: (1) DQN with the predictor trained on the original (non-augmented) dataset, (2) randomized sensing with an age-based predictor, (3) randomized sensing without an age-based predictor, and (4) the upper bound achieved with full sensing. 

\begin{figure}[t!]
\subfigure[Top-$1$ Accuracy] 
{
\includegraphics[width=0.22\textwidth]{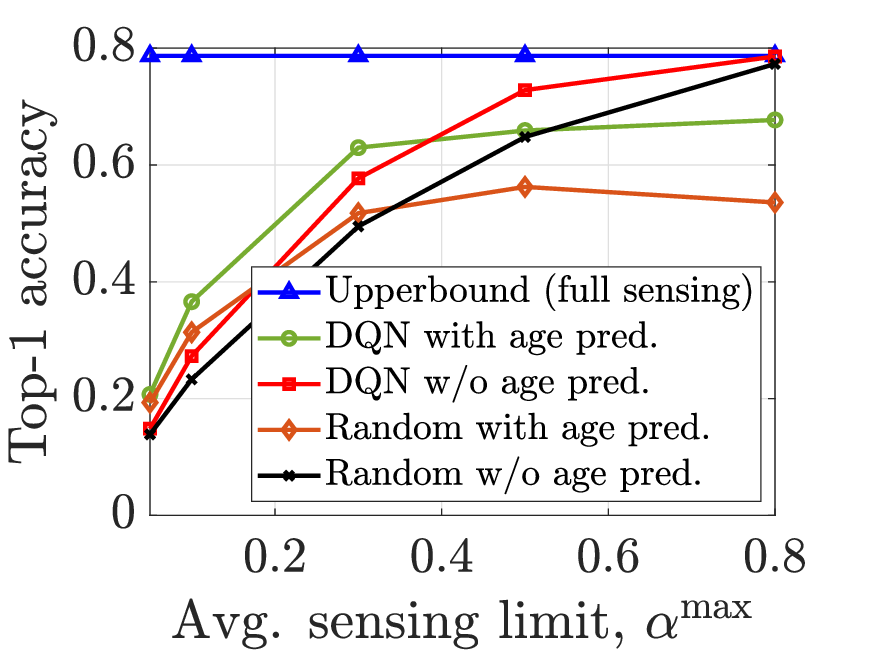}
\label{subfig_prf_top1}
}
\subfigure[Top-$3$ Accuracy]{
\includegraphics[width=0.22\textwidth]{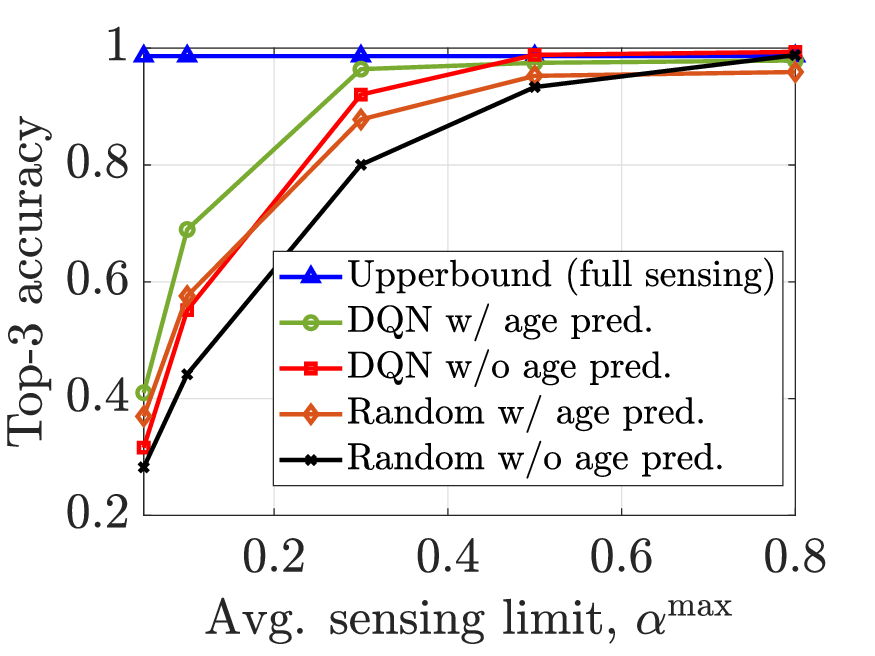}
\label{subfig_prf_top3}
}
\caption{Top-$k$ inference accuracy comparison for different sensing and prediction methods under varying sensing budgets $\alpha^{\max}$ for fixed age limit $N=5$ and parameter $V=[1~10~10~100~100]$ corresponding respectively to the points on the x-axis. }
    \label{fig_prf_com}
\end{figure}

Figure~\ref{fig_prf_com} illustrates the top-$1$ and top-$3$ inference accuracies under varying sensing budgets. The results show that the proposed method consistently outperforms all three baselines, particularly when the sensing budget is small. However, the performance of the age-based predictor in terms of top-$1$ accuracy decreases when the sensing budget exceeds $0.5$, which can be attributed to the effect of the age limit. As observed in Fig.~\ref{fig_age_limit}, for $\alpha^{\max}=0.5$, an age limit of $N=5$ becomes unnecessarily large, leading to a performance degradation due to the inclusion of overly stale samples in the training data.

\begin{figure}
    \centering
    \includegraphics[width=0.5\linewidth]{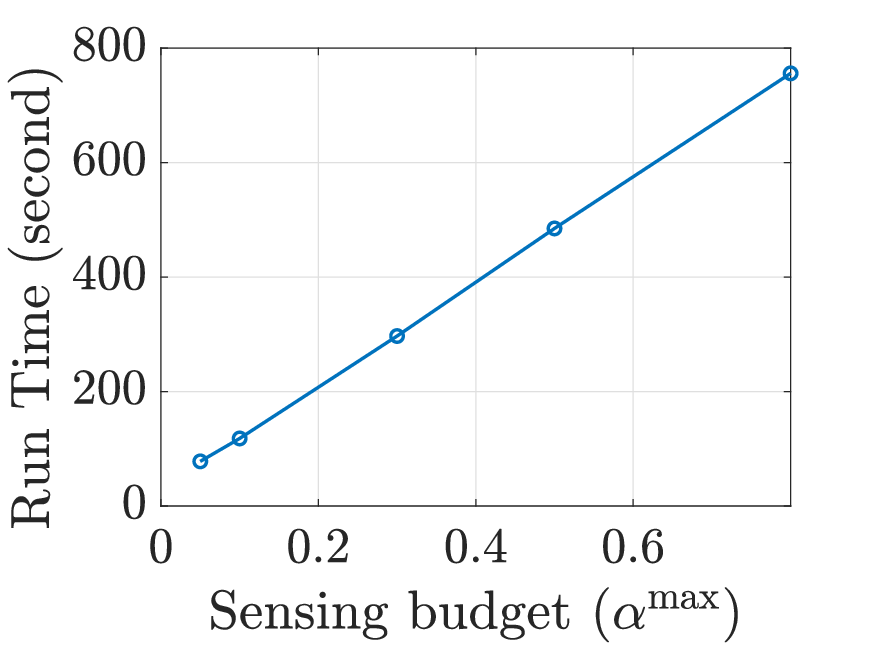}
    \caption{The inference runtime versus the sensing budget~$\alpha^{\max}$.}
    \label{fig_run}
    \vspace{-1 em}
\end{figure}

Finally, Fig.~\ref{fig_run} illustrates the inference runtime as a function of the sensing budget $\alpha^{\max}$. The runtime increases significantly as $\alpha^{\max}$ grows, confirming that executing more sensing operations and acquiring fresh data introduces higher computational complexity and consequently longer runtime. This highlights the importance of designing limited-sensing strategies for real-time applications. 
\section{Conclusions }\label{sec_conl}
We investigated the impact of sensing constraints on the inference performance of multimodal beam predictors using both position and image data. We proposed a joint sensing and beam prediction framework that integrates  DQN with a DNN-based beam predictor, where the age of information is incorporated into both the sensing policy and the training process. Lyapunov optimization was employed to enforce the average sensing constraint through a reward function.

Results show that incorporating data freshness during training improves inference performance under strictly limited sensing budgets. However, with relaxed sensing constraints, \textit{large} age limits in training may degrade performance--particularly top-$1$ accuracy--because using outdated data more often in the training complicates beam prediction, even though during inference, then more fresh samples will be available. Future work will consider distributed sensing across diverse datasets and deployment scenarios and will explore additional sensing~modalities.

 \section*{Acknowledgement}
This work was supported by the Research Council of Finland through 6G Flagship Program (no. 369116) and projects  DIRECTION (no. 354901) and DYNAMICS (no. 367702).
\vspace{-1 em}
\bibliographystyle{ieeetr}
\bibliography{Bib_References/conf_short,
Bib_References/IEEEabrv,
Bib_References/Bibliography, Bib_References/multimodalsensing_Bio}

\end{document}